\documentclass[aps,prb,twocolumn,superscriptaddress,showpacs,letter]{revtex4}
\usepackage{graphicx}
\usepackage{times}
\newcommand{\mb}[1]{\mathbf{#1}}
\newcommand{\mr}[1]{\mathrm{#1}}
\newcommand{\DN}{\delta N}
\newcommand{\De}{\delta\varepsilon}

\newcommand{\eps}{\varepsilon}
\newcommand{\DVp}{\delta U_p}
\newcommand{\Vg}{U_\mr{g}}
\newcommand{\Vp}{U_\mr{p}}
\newcommand{\Sz}{S_\mr{z}}
\newcommand{\zd}{z_\mr{d}}
\newcommand{\mean}[1]{\left\langle#1\right\rangle}
\newcommand{\ith}{i^{\text{th}} }
\newcommand{\Nth}{N^{\text{th}} }

\begin{document}
\title{Scrambling and Gate-Induced Fluctuations in Realistic Quantum Dots}

\author{Hong Jiang}
\thanks{Current address: Institut f\"ur Theoretische Physik,
  J.W.Goethe-Universit\"at, Frankfurt am Main, Germany} 
\affiliation{Department of Chemistry, Duke University, Durham, North
  Carolina 27708-0354} 
\affiliation{Department of Physics, Duke University, Durham, North
  Carolina 27708-0305}

\author{Denis Ullmo}
\thanks{Permanent address: Laboratoire de Physique Th\'eorique et
  Mod\`eles Statistiques (LPTMS), 91405 Orsay Cedex, France}  
\affiliation{Department of Physics, Duke University, Durham, North
  Carolina 27708-0305} 

\author{Weitao Yang}
\affiliation{Department of Chemistry, Duke University, Durham, North
  Carolina 27708-0354} 
\author{Harold U. Baranger}
\affiliation{Department of Physics, Duke University, Durham, North
  Carolina 27708-0305} 

\date{\today}

\begin{abstract}
We evaluate the magnitude of two important mesoscopic effects using a realistic model of typical quantum dots. 
``Scrambling'' and ``gate effect'' are defined as the change
in the single-particle spectrum due to added electrons or
gate-induced shape deformation, respectively. These two effects are
investigated systematically in both the self-consistent Kohn-Sham (KS)
theory and a Fermi liquid-like Strutinsky approach. 
We find that the genuine scrambling effect is small
because the potential here is smooth. In the KS theory, a key point is the implicit
inclusion of residual interactions in the spectrum; these dominate and make scrambling appear
larger. Finally, the gate effect is comparable in the two cases and, while
small, is able to cause gate-induced spin transitions.
\end{abstract}
\maketitle

\section{Introduction}
\begin{figure}[b]
\includegraphics[width=2.0in,clip]{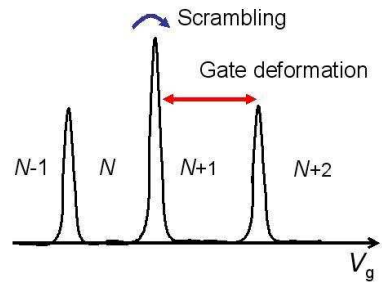}
\caption{\label{fig1} (Color online) Illustration of scrambling and gate
effects in Coulomb blockade conductance fluctuations. 
}
\end{figure}

An important way to characterize quantum dots
(QDs),\cite{Kouwenhoven97, Alhassid00RMP,Aleiner02} the simplest
artificial nano-structure with electrons quantized in all three
dimensions, is by the parametric evolution of their properties. The
most common external parameter is magnetic field because of its
flexibility of tuning,\cite{Kouwenhoven97} but other parameters are
also used. Here we are concerned with the effect of changing the
electron number $N$ or the external gate voltage $\Vg$, referred to as
the \textit{scrambling} and \textit{gate} effects, respectively, in
Coulomb blockade (CB) experiments. 
\cite{Sivan96,Stewart97,Patel98,Maurer99,Luscher01,Lindemann02,Patel98b}
The most striking feature of the CB regime is sharp peaks in the
conductance through the quantum dot as a function of gate voltage. As
shown in Fig.~\ref{fig1}, at each conductance peak, the number of
electrons residing in the dot changes by one; across a peak spacing,
the gate voltage changes to bring another electron into the dot, and
deforms the confining potential in the meantime. The scrambling and
gate-induced shape deformation effects were both
introduced\cite{Blanter97,Vallejos98} in connection with experiments
on the spacing between CB conductance
peaks,\cite{Sivan96,Patel98,Maurer99,Luscher01,Lindemann02} and have
also been used to interpret CB peak height
correlations.\cite{Patel98b}

The scrambling and gate effects can both be quantified through the
variation in the single-particle spectrum of the system,
$\{\varepsilon_i\}$. Since electron-electron interactions are
important for quantum dots in the Coulomb blockade regime, one must
clearly consider the effect of such interactions on the
single-particle spectrum. Here we evaluate the scrambling and gate
effects using both density functional theory and Thomas-Fermi
calculations for realistic geometries of quantum dots. We address two
main issues:

First, while the magnitude of the scrambling and gate effects has been
estimated for hard-wall quantum dots coupled to a large gate (one
which deforms the entire dot),\cite{Ullmo01b,Usaj02} experimental
quantum dots have, of course, smooth confining potentials, and are
typically deformed with a narrow ``plunger'' gate. We evaluate these
experimental features using our realistic model of quantum dots,
showing that they influence the magnitude of the scrambling and gate
effects strongly.

Second, what ``single-particle spectrum'' should one use in evaluating
the scrambling and gate effects? Roughly there are two types of
single-particle spectra that can be defined in an interacting system.
The first is a spectrum from a self-consistent mean field theory such
as Hartree-Fock (HF)\cite{SzaboOstlund} or Kohn-Sham spin-density
functional theory (KS-SDFT).\cite{ParrYang89} The second is the
spectrum of a reference Hamiltonian which contains the interactions
only at a smooth (classical-like) level.  The most natural choice is
the eigen-spectrum of the effective potential calculated from
Thomas-Fermi (TF) theory; this constitutes the Strutinsky approach
(S-TF).\cite{Ullmo01a,Ullmo04}
The difference between these two types of spectra is familiar, for instance, from discussions of the meaning of the eigenvalues in the self-consistent approach: Recall that the self-consistent eigenvalue is related to the energy for removing an electron from that level and that a sum over such eigenvalues double counts the interaction energy among those electrons,\cite{Messiah58,ParrYang89} neither of which is true for the eigen-spectrum in the reference Hamiltonian approach.

The meaning and magnitude of the scrambling and gate effects depend on
which type of single-particle spectrum is used. We emphasize that this
is \textit{not} a question of which approach is the more accurate, but
rather of what part of the fluctuations of the total energy is
assigned to these effects. For instance, when using a reference
potential as in S-TF, the fluctuations as a parameter changes
associated with interactions are separated into two distinct parts.
The first comes from changes in the single-particle energies as the
smooth Thomas-Fermi potential varies -- we consider these the
``genuine'' scrambling and gate effects. The second contribution
involves the screened Coulomb interaction between mesoscopic density
fluctuations. This ``residual interaction'' term corresponds to the
weak interaction between Landau quasi-particles in a Fermi liquid
picture.  In a self-consistent approach such as KS-SDFT, they appear
in the self-consistent energies $\varepsilon_i^{\rm KS}(N)$; thus
parametric evolution of the KS levels involves both genuine scrambling
and residual interactions.

The paper is organized as followings. In the next section, we outline
briefly the two theoretical approaches, i.e. the Kohn-Sham method and
Fermi-liquid like Strutinsky approximation scheme. In Section III, we describe the 2D realistic quantum dot model used for the study the scrambling and gate effects. The main results of this study are presented and analyzed in Section IV. 

\section{Theory}

In both the KS-SDFT and S-TF approaches, for a system with $N$
electrons and total spin $\Sz$, one solves a Schr\"odinger equation
with a spin-dependent potential,
   \begin{equation} 
\big\lbrace -\frac{1}{2}\nabla ^{2}+U^\sigma(\mb{r}) \big\rbrace \psi
_{i}^{\sigma}(\mb{r})=\varepsilon _{i}^{\sigma } \psi    _{i}^{\sigma
}(\mb{r}) \label{eq:KS}  
   \end{equation}
\vspace*{-0.15in}
   \begin{equation} 
U^\sigma(\mb{r}) = U_\mathrm{ext}(\mb{r}) +
  U_{H}(\mb{r};[n]) + U_{\mathrm{xc}}^{\sigma }(\mb{r};[n^{\alpha
  },n^{\beta }]) 
       \label{eq:VKS} 
   \end{equation} 
where the total potential is the sum of the external, Hartree, and
exchange-correlation contributions, and $\sigma=\alpha,\beta$ denotes spin-up and down, respectively. In KS-SDFT,\cite{ParrYang89} one
knows that the potentials are functionals of the spin densities
$n^{\sigma }(\mb{r})  \!=\!  \sum_{i}^{N^\sigma} \left| \psi
_{i}^{\sigma}(\mb{r})\right|^{2}$, which are solved self-consistently
under the constraint $\int n^{\sigma} (\mb{r}) d\mb{r}  \!=\!  N^{\sigma}$
with $ N^\alpha \!=\! (N+2\Sz)/2$ and $N^\beta  \!=\!  (N-2\Sz)/2$. In analogy to
the Koopmans's theorem in Hartree-Fock theory, it has been proved that
the highest occupied Kohn-Sham orbital energy is identical to the
chemical potential if the {\em exact} exchange-correlation potential
is used.\cite{Almbladh85} This association provides a physical meaning to the self-consisent eigenvalues which can be contrasted with that of the reference Hamiltonian eigenvalues.

In the S-TF approach, the basic idea is to start from a smooth
semiclassical approximation, i.e. the Thomas-Fermi theory, and
introduce quantum interference by considering, first, single-particle
corrections and, then, the effect of screened interactions between the
oscillating part of the electron density. In this case, then, one uses
the TF potential in Eq.~(\ref{eq:KS}). From the resulting
$\varepsilon_i^\sigma$ and $\psi_i^\sigma$ one obtains an
approximation to the KS-SDFT total energy valid up through
second-order in the oscillating density, $ n^{\sigma }_{\rm
osc}(\mb{r}) \equiv n^{\sigma }(\mb{r}) -
n_{\text{TF}}(\mb{r})$.\cite{Ullmo01a,Ullmo04} In this approximation,
the relation between the KS-SDFT and Thomas Fermi single-particle
levels can be expressed as\cite{Ullmo04}
\begin{equation}
   \varepsilon_{i,\sigma}^{\rm KS} \approx  
   \varepsilon_{i,\sigma}^{\rm TF} + \delta \varepsilon^{\rm RI}_{i,\sigma}
   \label{eq:epsilons}
\end{equation}
where the residual interaction terms
\begin{equation} \label{eq:RI}
\delta \varepsilon^{\rm RI}_{i,\sigma} = \sum_{\sigma'} \int d\mb{r} d\mb{r'}
|\psi_i^\sigma(\mb{r})|^2 V^{\sigma,\sigma'}_{\rm scr}(\mb{r},\mb{r'})
n^{\sigma' }_{\rm osc}(\mb{r'}) 
\nonumber   
\end{equation}
corresponds to the interaction of $\psi_i^\sigma$ with the density
ripples due to interferences.  The screened interaction potential
$V^{\sigma,\sigma'}_{\rm scr}(\mb{r},\mb{r'})$ is expressed 
explicitly  in terms of the  functional derivatives of
$U^\sigma(\mb{r};[n^{\alpha}])$.\cite{Ullmo04}  When for instance an electron is
added into the quantum dot,  both the genuine scrambling effect (i.e.\
the variations of the Thomas Fermi levels) {\em and} the residual
interaction terms Eq.~(\ref{eq:RI}) will affect the KS-spectrum.

In this study we evaluate the scrambling and gate effects in both KS
and S-TF approaches. For the S-TF case, the spin-dependence of the
potential has little effect on its spectrum other than a constant
shift, so we use the spin-independent Thomas-Fermi potential. For KS,
we first solve the full spin-dependent KS equations, but calculate the
scrambling and gate effects only from $\alpha$ orbital energies. Only
minimal spin states, $\Sz \!=\! 0$ for even $N$ and $\Sz \!=\! 1/2$ for odd $N$,
are considered. The spin indices will therefore be dropped in the
remainder of the paper. The numerical methods that are used to solve KS and TF equations are described in details in Refs.\onlinecite{Jiang03b,Jiang04b}, respectively. 

\section{Model system}
The model system we use for investigating gate and scrambling effects
is a realistic 2D lateral quantum dot.\cite{Jiang04a} The electrons
are at the heterointerface a distance $z_d$ below the surface of the
heterostructure. For the electrostatic potential, we use the mid-gap
pinning model for the boundary condition at GaAs free surface:
\cite{Nixon90, Davies94, Davies95} We impose Dirichlet boundary
conditions on the top surface and Neumann conditions in infinity in
all other directions,\cite{Jiang04a,footnote-boundary} allowing the
external potential to be calculated from
\begin{equation}U_\mathrm{ext}(\mathbf{r})=\frac{1}{2\pi}\int\!\!
  d\mathbf{r}'\Vg(\mathbf{r}')
  \frac{\zd}{(|\mathbf{r}-\mathbf{r}'|^2+\zd^2)^{3/2}}+
  U_\mathrm{QW}(\zd),\label{eq:Vconf}
\end{equation} 
where $\Vg(\mb{r})$ is the electrostatic potential on the top gate
surface,\cite{Davies95} and $U_\mathrm{QW}(z)$ is the confining
potential in the growth direction due to the quantum well structure
from which the quantum dot is fabricated. In addition, the Hartree
potential has an image term due to the coupling with the top surface,
\cite{Nixon90, footnote-boundary}
 \begin{equation}
U_\mathrm{H}(\mathbf{r};[n])=\int\!
d\mathbf{r}'n(\mathbf{r}') \big[\frac{1}{|\mathbf{r}-\mathbf{r}'|}
-\frac{1}{(|\mathbf{r}-\mathbf{r}'|^2+4\zd^2)^{1/2}} \big] \,.
\end{equation}
For a complete description of our treatment of the electrostatic
potential see Ref.\ \ \onlinecite{Jiang04a}.

\begin{figure}
\includegraphics[width=1.6in,clip]{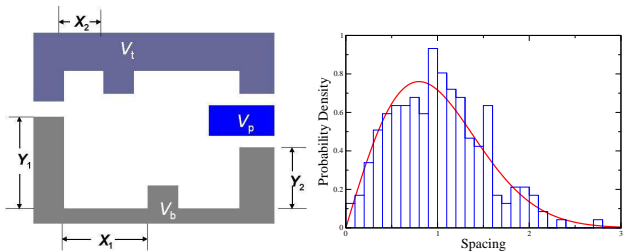}
\includegraphics[width=1.6in,clip]{GSE-nnsd.eps}
\caption{\label{fig:shape} (Color online) Left panel: Schematic of the
shape of the top confining gate used in this study. Negative voltages
are imposed on the shaded region, depleting the electrons underneath
so that the motion of electrons is confined to a small region. To
obtain enough statistics, 24 different irregular confining shapes are
generated by taking different values of $X_1$, $X_2$, $Y_1$ and $Y_2$.
Right panel: The nearest neighbor spacing distribution of the
single-particle levels calculated in the Thomas-Fermi effective
potential (histogram) compared to the Wigner surmise distribution
(line).  }
\end{figure}

The shape of the top confining gate used here is shown in
Fig.\ref{fig:shape}; it is designed to model typical irregular dots
investigated experimentally.\cite{Patel98} Negative voltages are
imposed on the top ($U_\mr{t}$), bottom ($U_\mr{b}$), and plunger
($\Vp$) electrode gates; the electron number in the dot is controlled
by $\Vp$. The single-particle dynamics of the system is expected to be
chaotic, which is confirmed by the agreement between the
nearest-neighboring spacing distribution of the single-particle levels
and the Wigner surmise distribution.\cite{Bohigas90}

We now introduce quantities to characterize the scrambling and gate
effects. For the scrambling effect, we define
\begin{equation}
\De_i(\DN) \equiv \eps_i(N^{(0)}+\DN)-\eps_i(N^{(0)}).
\end{equation}
where $\eps_i(N)$ is the $\ith$ single-particle energy in either the
TF or KS effective potential with $N$ electrons. Scrambling is
quantified by the magnitude of fluctuations in $\De_i(\DN) $:
\begin{equation}
\sigma_\mr{s}(\DN) \equiv \sigma\left\lbrace
\frac{\De_i(\DN)-\left\langle \De_i(\DN) \right\rangle
}{\Delta}\right\rbrace \label{eq:sigma_s}
\end{equation}
where $\left\langle \De_i(\DN) \right\rangle $ denotes a linear fit of
$ \De_i(\DN) $ as a function of $i$, and $\Delta$ is the mean level
spacing.  The gate effect is similarly characterized as
\begin{equation}
\sigma_\mr{g}(\DVp) \equiv \sigma \left\lbrace 
\frac{\De_i(\DVp)-\mean{\De_i(\DVp)}}{\Delta}\right\rbrace
\label{eq:sigma_g}
\end{equation}
with $\De_i(\DVp) \equiv \eps_i(U_p^{(0)}+\DVp)-\eps_i(U_p^{(0)})$,
and $\mean{\De_i(\DVp)}$ its linear fitting. It is more convenient to
write $\sigma_\mr{g}$ as a function of $\delta N^*\equiv \DVp/
\mean{\delta U^0}$, where $\mean{\delta U^0}$ is the average
conductance peak spacing.  $\delta N^*$ can be regarded as the induced
electron number due to a change of the gate voltage.

\section{Results and Discussion}
\label{sec:discussion}
The scrambling and gate effects are mixed together in CB conductance
peak spacings. It is desirable, however, to first study them
separately; while the separation of the two effects is difficult to
implement experimentally and requires sophisticated
design,\cite{Patel98} it is straightforward in numerical
investigations. For the scrambling effect, we fix the external
confining potential and calculate the TF and KS single-particle
spectra at each $N\in[50,70]$. For the gate effect, we fix $N \!=\! 70$ and
$\Sz \!=\! 0$, and scan the plunger gate voltage for $\delta N^*$ up to
about $20$. Statistics are obtained by averaging over different levels
and 24 confining gate shapes.

\begin{figure}[tb]
\includegraphics[width=2.8in,clip]{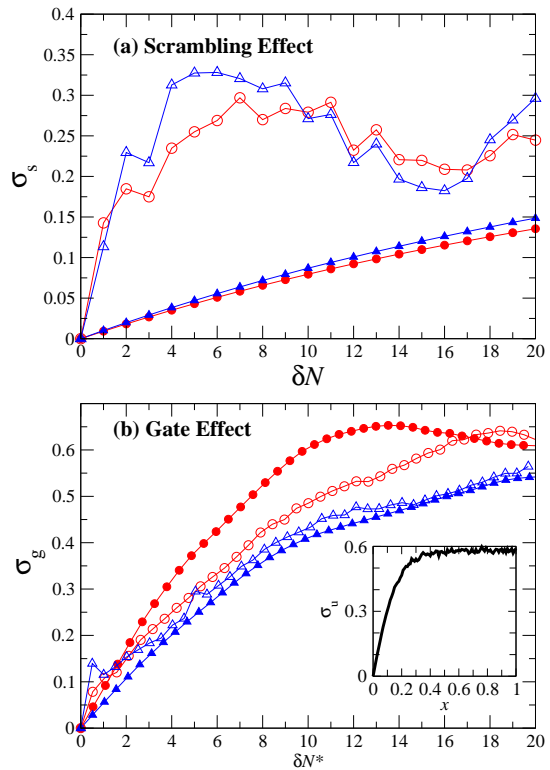}
\caption{\label{fig:sigma} 
(Color online) Scrambling magnitude as a function of $\delta
N$ (upper) and 
gate fluctuation as a function of $\delta N^*$ (lower) from the TF
(filled) or KS (empty) single particle spectra with $z_\mr{d} \!=\! 15$ nm
(circles) or $50$ nm (triangles). Inset: The result of a universal
random matrix model.  }
\end{figure}

Fig.~\ref{fig:sigma}(a) shows the scrambling effect calculated from
both TF and KS single-particle spectra. A remarkable feature is
the large difference between TF and KS results. $\sigma_\mr{s}$ from
TF spectra increases smoothly as a function of $\delta N$, but is
smaller than 0.15 even at $\delta
N \!=\! 20$.\cite{footnote-sum-contribution} In contrast, $\sigma_\mr{s}$
from KS spectra increases rapidly at first, and saturates for $\delta
N>4 $.  For $\delta N>8 $, $\sigma_\mr{s}$ shows some modulations, the
reason for which is not yet clear. The effect of the dot depth,
$\zd \!=\! 15$~nm vs. $50$~nm, is quite small. \textit{The fact that the KS
results are much larger than TF ones shows that the residual
interaction effects included in the self-consistent KS energies [Eq.~(\ref{eq:epsilons})]
dominate the genuine scrambling evaluated with TF.}

Fig.~\ref{fig:sigma}(b) shows the gate effect calculated from both TF
and KS single-particle spectra. The gate effect in TF and KS spectra
are qualitatively the same; in both cases $\sigma_\mr{g}$ first
increases and then saturates at some large $\DN^*$ (about $10$ for
$\zd \!=\! 15$ nm in the TF case). The saturated values of $\sigma_\mr{g}$
are about 0.6, larger than that of $\sigma_\mr{s}$(KS).  The gate
effect is quite sensitive to the depth of the dot, especially for the
TF case; it is larger for shallow dots, as expected because the gate
becomes sharper and better defined. Finally, note that the gate effect
is about one order of magnitude larger than the TF scrambling effect.

To further understand these results, we model the parametric evolution
of the single-particle spectrum by using a random matrix Gaussian
process,\cite{Alhassid99}
\begin{equation}
H(x)=\cos (x\pi/2) H_1+\sin(x\pi/2) H_2 
\end{equation}
where $H_1$ and $H_2$ are independent random matrices belonging to the
Gaussian orthogonal ensemble (GOE).\cite{Bohigas90,Alhassid00RMP} We
define $\sigma_\mr{u}(x)$ similarly to Eqs. (\ref{eq:sigma_s}) or
(\ref{eq:sigma_g}) to characterize the change in the single-particle
spectrum of $H(x)$ due to the variation of $x$. The inset of
Fig.~\ref{fig:sigma}(b) shows $\sigma_\mr{u}$ vs. $x$ obtained from
500 realizations of $30\times30$ random matrices. $\sigma_u$ saturates
at about $x \!=\! 0.4$, and the saturated value is about $0.6$ in agreement
with that of $\sigma_\mr{g}$. We notice that the saturation behavior
of $\sigma_\mr{s}$ from KS levels is very different from that of
$\sigma_\mr{u}$; this is because the KS single-particle levels contain
implicitly some interaction effects so that their variation with $N$
is not a simple Gaussian process.

The agreement between the KS and TF results for the gate effect in Fig.~\ref{fig:sigma}(b) is particularly striking when compared to the sharply different magnitudes for the scrambling effect in Fig.~\ref{fig:sigma}(a).
The underlying reason for this difference between the gate and scrambling effects is the extra electron added in the case of scrambling: First, the contribution to the fluctuations of $\varepsilon_{i,\sigma}^{\rm KS}$ 
by most of the residual interactions is small because of the small change in wave functions caused by the additional potential. Second, the contribution by residual interactions involving the added electron $j$ are, however, much larger: the fluctuations here involve the deviation of 
$|\psi_j^\sigma(\mb{r})|^2 $ from the smooth density rather than the small change in an already filled level. Thus the fluctuation of the residual interaction contribution to $\varepsilon_{i,\sigma}^{\rm KS}$ is substantially stronger in the case of scrambling than for the gate effect.

We now turn to actual spacings between CB conductance peaks and
calculate the scrambling and gate effects. The position for the $\Nth$
peak, $U_g^{(N)}$, at which the energies for $N-1$ and $N$ electrons
are equal, is determined by $\mu(N,U_g) \equiv
E_\mr{KS}(N,U_g)-E_\mr{KS}(N-1,U_g)  \!=\! 0 $ if chemical potentials in the
leads are taken to be zero. From the TF spectrum at this $U_g$, the
scrambling and gate effects are the standard deviation of
$\eps_i(N,U_g^{(N)})-\eps_i(N-1,U_g^{(N)})$ and
$\eps_i(N,U_g^{(N+1)})-\eps_i(N,U_g^{(N)})$, respectively. For
$\zd \!=\! 15$~nm, the former is equal to $0.009 \Delta$, and the latter
$0.07 \Delta $, using the same confining gate shapes and parameter
ranges as above. Note the good agreement with the results in
Fig.~\ref{fig:sigma} for $\delta N \!=\! 1$ or $\delta N^* \!=\! 1$.

Comparison with earlier evaluations of these effects yields important
insights. We start with the scrambling effect; in particular, using
the expressions derived in Refs.~\onlinecite{Ullmo01b,Usaj02} for
scrambling associated with the Thomas-Fermi spectrum -- which we think
of as the ``genuine''scrambling effect -- yields $\sigma^{\rm
pred}_s(\delta N\!=\!1) \simeq 0.06$ for a dot with $N \!=\! 70$ electrons.
This prediction is six times larger than the value obtained here.
There are two main differences between the earlier situation and ours:
the confining potential here is smooth while it was assumed to be
hard-wall in Refs.~\onlinecite{Ullmo01b,Usaj02}, and we effectively
have a top gate across the whole dot because of our boundary
condition. The insensitivity of the scrambling magnitude to the
spacing $z_d$ suggests that the top gate has little effect. {\em We
therefore conclude that there is significantly less scrambling in a
smooth confining potential than in a hard wall dot}.

Further support for this conclusion comes from the change in potential
upon adding an electron: the hard-wall gives rise to a square-root
singularity in this quantity;\cite{Blanter97} the absence of a
singularity in the smooth case naturally leads to a weaker effect. As
a consequence, the scrambling in the KS spectra, which contains both
genuine scrambling and some residual interaction fluctuations [see
Eq.~(\ref{eq:epsilons})], appears to be entirely dominated by the latter.
Evaluating the magnitude of these residual interaction fluctuations as
an extra electron is added, following the semiclassical random plane
wave approach in Refs.~\onlinecite{Ullmo01b,Usaj02}, gives $0.18
\Delta$ for $N \!=\! 70$. Taking into account that one is not very far into
the semiclassical regime, this is quite compatible with the value
$\sigma_s(\delta N^*\!=\!1) \!\approx\! 0.13$ computed for the Kohn-Sham
spectra [Fig.~\ref{fig:sigma}(a)].

For diffusive transport in a weak disordered potential, scrambling has
been studied using the statistics of single-particle wavefunctions in
that case.\cite{Alhassid00RMP} In that context, scrambling grows linearly
with $\delta N$ while residual interaction effects grow as $(\delta
N)^{1/2}$. It is interesting to note that our data in
Fig.~\ref{fig:sigma}(a) for a ballistic dot show roughly the same
behavior.

With regard to the gate effect, it was argued that a ``generic'' gate
should have the same effect as TF scrambling \cite{Usaj02}. For the
gate considered here, we however observe a significant difference.  A
``generic'' gate is one which couples approximately uniformly to the
dot -- a back gate, for instance.  The opposite extreme is a gate
coupled very locally to a point in the dot, thus producing a rank one
perturbation.  Such perturbations are known to completely decorrelate
the spectra for the phase shift $\pi/2$ necessary to add an extra
electron.\cite{Aleiner98} Here we see that lateral plunger gates are
an intermediate case: they produce fluctuations which are
significantly stronger than the scrambling effect but remain moderate
on the scale of the mean level spacing.

The gate effect here is nevertheless strong enough to produce spin
transitions. Recently Kogan, et al. reported experimentally a
singlet-triplet transition in zero magnetic field driven by changing
the confining potential.\cite{Kogan03} In our calculations, we also
observe spin transitions caused by gate-induced shape deformation.
The probability of a transition -- the fraction of peak spacings in
which the ground state spin flips as $\Vp$ changes -- is $5.8\%$ for
$\zd \!=$ $15$~nm and $4.6\%$ for $\zd \!=$ $50$~nm.  In addition to
the simple singlet-triplet transition, different patterns appeared in
our calculations.  In many cases, the spin flips are paired: in a
single spacing interval, the spin changes from one value to another
and then back again, which agrees with the general picture connecting
spin transitions to avoided-crossings caused by
shape-deformation.\cite{Baranger00} In some rare cases,
triple-transitions are observed, such as $\Sz \!=\! 2\!\rightarrow\! 1$
$\!\rightarrow\! 0$ $ \!\rightarrow\! 1$ or $3/2\!\rightarrow\! 1/2$
$\!\rightarrow\! 3/2 $ $\!\rightarrow\! 1/2$.  In other cases, the
spin transition is unpaired, presumably because pairing due to an
avoided-crossing is destroyed by a change in electron number.

Several experiments have traced the correlation between ground states
or excited states as the number of electrons
changes,\cite{Stewart97,Patel98b,Maurer99,Luscher01} and often see a
surprising degree of correlation. The relatively weak scrambling and
gate effects that we find offer a way to understand these results.

To conclude, we have investigated the scrambling and gate effects for
two different one-particle spectra -- self-consistent Kohn-Sham and
Thomas-Fermi -- of a realistic model quantum dot.  Our main findings
are: (1) The genuine scrambling effect -- the one associated with the
Thomas-Fermi spectra -- is {\em significantly smaller} for the smooth
potential considered here than in earlier work using hard wall
confinement.  (2)~As a consequence, scrambling for the Kohn-Sham
spectra, which involves both genuine scrambling and residual
interactions, is entirely dominated by residual interactions. Its
magnitude agrees with those from a random plane wave model of the wave
functions.  (3) Finally, fluctuations caused by the gate are similar
in magnitude for the two spectra. A lateral plunger gate causes
significantly larger fluctuations than those caused by scrambling,
contrary to the case of a uniform ``generic'' gate.  The magnitude of
the gate effect remains nevertheless moderate compared to the mean
level spacing.

\begin{acknowledgments}
We thank G. Usaj for valuable comments. This work was supported in
part by NSF Grant No. DMR-0103003.
\end{acknowledgments}


\begin{thebibliography}{35}
\bibitem[{foo({\natexlab{a}})}]{footnote-boundary}
\bibinfo{note}{The underlying physics of the pinned-level model is to assume
  that electrons can freely move between GaAs surface states and the 2D
  electron gas, so that the chemical potential is pinned by the surface states
  in exposed regions. Indeed the validity of the pinned-level model, though
  well established for GaAs/AlGaAs devices in equilibrium at finite
  temperature, is still an open question for lateral quantum dots at near-zero
  temperature (see, for example, recent experimental studies on this issue in
  Ref. \onlinecite{Kawaharazuka01}). Here it is worth mentioning that another
  approach for realistic modeling of quantum dots is to assume the same
  boundary conditions in a finite volume instead of at infinity as we did (see,
  for example, Ref. \onlinecite{Zhang04}). Since the top confining metal gates
  are usually on the scale of microns, much larger than the quantum dot itself,
  the difference between these two treatments should be minor. This is actually
  supported by the fact that, as stated in Sec.~\ref{sec:discussion}, the
  effect of the dot depth, $\zd \!=\! 15$ nm compared with $50$ nm, is quite
  small. With the assumed boundary condition, the image term in the Hartree
  potential arises naturally by taking advantage of the linearity of Poisson
  equation.}
\bibitem[{foo({\natexlab{b}})}]{footnote-sum-contribution}
\bibinfo{note}{In contributing to the conductance peak spacing, the scrambling
  effect contains two parts:\cite{Ullmo01b} the contribution from the top level
  -- the quantity discussed here -- and that from the sum over all other
  occupied levels. In our numerical calculations, we confirm the expectation
  that the second term is small compared to the first.}
\end{thebibliography}

\end{document}